# Polarons, free charge localisation and effective dielectric permittivity in oxides

Mario Maglione
ICMCB-CNRS, University of Bordeaux 1
87, Av Dr Schweitzer 33608 Pessac France
maglione@icmcb-bordeaux.cnrs.fr

Abstract

This review will deal with several types of free charge localisation in oxides and their consequences on the effective dielectric spectra of such materials. The first one is the polaronic localisation at the unit cell scale on residual impurities in ferroelectric networks. The second one is the collective localisation of free charge at macroscopic interfaces like surfaces, electrodes and grain boundaries in ceramics.

Polarons have been observed in many oxide perovskites mostly when cations having several stable electronic configurations are present. In manganites, the density of such polarons is so high as to drive a net lattice of interacting polarons. On the other hand, in ferroelectric materials like $BaTiO_3$ and $LiNbO_3$, the density of polarons is usually very small but they can influence strongly the macroscopic conductivity. The contribution of such polarons to the dielectric spectra of ferroelectric materials is described. Even residual impurities as for example Iron can induce well defined anomalies at very low temperatures. This is mostly resulting from the interaction between localised polarons and the highly polarisable ferroelectric network in which they are embedded. The case of such residual polarons in $SrTiO_3$ will be described in more details, emphasizing the quantum polaron state at liquid helium temperatures.

Recently, several non-ferroelectric oxides have been shown to display giant effective dielectric permittivity. It is first shown that the frequency/temperature behaviour of such parameters is very similar in very different compounds (donor doped $BaTiO_3$, $CaCu_3Ti_4O_{12}$, $LuFe_2O_4$, Li doped NiO,…). This similarity calls for a common origin of the giant dielectric permittivity in these compounds. A space charge localisation at macroscopic interfaces can be the key for such extremely high dielectric permittivity.

# I. Introduction

Free charges may be generated in dielectric oxides by ionic substitution and/or creation of oxygen vacancies. In the case of ferroelectric perovskites such solid state chemistry processes have attracted a lot of interest and are still a matter of debate in the recent literature [1,Chan N. H., 1981],[2,Smyth D. M., 2000a], [3,Wollman M., 1994], [4,Waser., 1992], [5,Smyth D. M, 2002b], [6,Morrison F. D., 2001]). It is out of the scope of this chapter to actually face this question. However, we have to point out from the very beginning that charged point defects are present in oxides, whether unwanted or intentionally included. In this chapter, we will focus on the contribution of such defects to the dielectric properties of materials.

We will start from the case of low density charged defects resulting in electronic localisation at the unit cell length scale. In highly polarisable lattices like oxides perovskites, such localisation results in a lattice deformation and polarisation. Since such Polarons are carrying a charge, an electric dipole moment and an elastic energy, they can be probed by a large variety of experimental techniques. Local spectroscopies (ESR [7,Possenriede],) optical spectroscopies ( [8,Prosandeev,2002], [9,Trepakov, 2006]) and a coupling between them [10,Thiemann, 1991] are useful to probe the electronic state of polarons. Even being highly debated in the case of a high density of defects [11,Scott,2003], these techniques provide well defined signatures which can be compared to models based on semi-conductor picture of the host matrix. Conductivity[12,Eagles,….], drift mobility [13,Michel Calendini] and Hall experiments are the key ways to check the thermally activated motion of polarons. These tools are however restricted to the cases where the density of polarons is large. These cases will not be addressed in this chapter since a number of review papers already appeared 14.[Elliott, 1987], 15.[Mott ,1979]. We will rather focus on the cases where the density of such polarons is so small as to invalidate usual experiments probing mobile charges. In slightly doped oxidic $ABO_3$ we will describe dielectric spectroscopy which was able to probe very small thermally activated dielectric losses which were assigned to the dynamics of polarons. The correlation of such polarons with the high polarisability of the underlying lattice will be described. Intentionally substituted point defects in perovskites as well as unwanted residual impurities will be listed. In this latter case, particular attention will be paid on Fe substitution on the B octahedral site which is an unavoidable defect. Because of the two stable valence states of this cation, polaron hoping is well documented.

The second part of this chapter will deal with another case in which free charge localisation among ionic sites do contribute to the overall dielectric properties of materials, namely the large scale and collective localisation of space charges. Being well known in the field of semiconductors where free charges depletion at interfaces is the basis of all microelectronic devices [16,Sze], these is still a matter of debate in oxides. As in semiconductors, the heterovalent substitutions at ionic site in perovskites are to provide free charges which will then localise and increase artificially the dielectric permittivity. The first process for this is the creation of oxygen vacancies through a careful control of the reducing atmosphere to which bulk materials are submitted [17,Stumpe], [18,Waser], [19,Bidault]. This space charge effect is a very general trend of bulk ferroelectric annealed at high temperature [19,Bidault]. Because of the restricted geometry and of the sharp processing conditions, such space charges can be localised at room temperature in thin films [20,Dawber, Stolicnov]. The next way of generating space charges is the cationic heterovalent substitution. Couples like $Nb^{5+}/Ti^{4+}$, $Ba^{2+}/La^{3+}$, $Pb^{2+}/La^{3+}$, $Bi^{3+}/Pb^{2+}$ are among the most popular but this list is by no way complete. Even in pure lattices where heterovalent cations like Cu, Fe, Mn are present in a very ordered way in every unit cell, interfacial charge depletion can lead to giant dielectric properties ($CaCu_3Ti_4O_{12}$[21,CCTO Subramanian], $LuFe_2O_4$ [22,Ikeda], [23,Wu],

$AFe_{1/2}B_{1/2}O_3$. A=Ba, Sr, Ca; B=Nb, Ta, Sb [24,Raevski]. The differences between this effective dielectric behaviour and the really microscopic polarisability will be given.

**II Interplay between conductivity and dielectric permittivity in impedance experiments.**

In semi-conductors [25,Pollak 1961], [14,Elliott 1987] and ionic conductors [26,Ming],[27,MacDonald],[28,Hodge],[29,Macedo], free charges and lattice polarisability have contribution to the overall dielectric properties. The usual way to discriminate between these two contributions is to perform impedance spectroscopy on a broad frequency range [30,Jonscher, 1983] [31, Coehlo,1978]. To do so, one has to draw an equivalent circuit for the samples which, neglecting the inductance part is simply a parallel resistor-capacitor (RC) circuit. Using this, one can access from the complex impedance $Z^*(\omega) = Z'(\omega) - jZ''(\omega)$ to either the complex conductivity $\sigma^*(\omega) = \sigma_1(\omega) - j\sigma_2(\omega)$ or to the complex dielectric permittivity $\varepsilon^*(\omega) = \varepsilon_1(\omega) - j\varepsilon_2(\omega)$. It is to be pointed out that in many cases, both parameters can be computed from a single experiment. Figure 1 is showing the example of the simplest RC parallel circuit where both the resistance and the capacitance are frequency independent. On figure 1a, the so called Nyquist plot –Z" versus Z' has the well known semi circle shape from which R and C can be computed. Using the same R and C, one can plot $\varepsilon_2$ versus $\varepsilon_1$ which in this case is a straight line (figure 1b). To get a semi-circle or a full circle in this $\varepsilon_2/\varepsilon_1$ plane, one should have a frequency variation of R(ω) and C(ω). This striking difference between the Z"/Z' Nyquist plot and the $\varepsilon_2/\varepsilon_1$ Cole-Cole plot was already recognized back in 1941 by Cole and Cole[32,Cole-Cole,1941]. They showed that during a dielectric relaxation the impedance phase angle is constant which means that all points are coinciding in the Nyquist plane when they display a semi-circle in the Cole Cole plane. The resulting shape of $\varepsilon_1(\omega)$ and $\varepsilon_2(\omega)$ is schematically shown on figure 2 on a ultra-broad frequency range from $10^{-2}$Hz up to $10^{20}$Hz [31,Coehlo,1978]. One can distinguish between relaxations at frequencies lower than about $10^{10}$Hz and resonances above this. Relaxations result from the dynamical friction of electric dipoles with their surrounding while resonances stems from the restoring forces acting on charges. It is out of the scope of the present chapter to describe all of this processes which have been already reviewed long ago [30, Jonscher], [31,Coehlo]. For the purpose of looking for interaction between free charges and lattice polarisability, one should focus on the low frequency range of figure 2 in the sub-MHz frequency range $f<10^6$Hz. Again, in a very schematic way, $\varepsilon_1$ and $\varepsilon_2$ undergo dispersions called dielectric relaxation. When plotted in the Cole-Cole plane[32,Cole-Cole,1941], semi-circle like diagrams are recovered. On the assumption that the dipoles which lead to such relaxations are independent one from another, the analytical equation for dielectric relaxation is the Debye formula

$$\varepsilon^*(\omega) = \varepsilon_\infty + \frac{\varepsilon_s - \varepsilon_\infty}{1 + (i\omega\tau)^{1-\alpha}} \qquad (1)$$

The essential relaxation parameters are the relaxation time τ and the dispersion step $\varepsilon_s - \varepsilon_\infty$. The Cole-Cole exponent α accounts for the possible disorder among the relaxing dipoles. Whenever the conductivity cannot be neglected, a base line should be added to equation (1) which contribute mainly to the real part of the impedance Z' and thus to the imaginary part of the dielectric permittivity $\varepsilon_2(\omega)$. In terms of imaginary permittivity, this term is $\frac{\sigma_{dc}}{\omega}$ where $\sigma_{dc}$ is the static conductivity. However, to take into account the possible interaction of free charges with their underlying lattice, a complex conductivity may be included $\sigma^*(\omega) = \sigma_1(\omega) - j\sigma_2(\omega)$. The imaginary conductivity (which is much smaller than the real one) will contribute to the real part of the permittivity and vice versa. Moreover, disorder effects in the conductivity are taken into account with a frequency exponent smaller than 1 in the conductivity term. As a result, the dielectric dispersion in the presence of disordered dynamical conductivity reads:

$$\varepsilon^*(\omega) = \varepsilon_1(\omega) - j\varepsilon_2(\omega) = \varepsilon_\infty + \frac{\sigma_1}{\omega^s} + j\frac{\sigma_2}{\omega^s} + \frac{\varepsilon_s - \varepsilon_\infty}{1+(i\omega\tau)^{1-\alpha}} \quad (2)$$

In the following, the real and imaginary parts of the dielectric permittivity will be fitted separately with $\sigma_1$, $\sigma_2$, s, $\varepsilon_s - \varepsilon_\infty$, $\tau$ and $\alpha$ as fitting parameters. For such fits to be reliable, the operating frequency range should be large enough (4 decades at least) and the conductivity and relaxation processes should be away one from another as for the fitting parameters to be uncorrelated. On changing the sample temperature, the relative strength of the conductivity low frequency tail and of the relaxation at higher frequency may be discriminated. Equation (2) is quite powerful since a single experiment with the same sample in the same sample holder can provide both the conductivity and the dielectric relaxation step. As we will see in the following, this is very useful when looking for contributions of the conductivity in the effective dielectric properties. We stress that such an interrelation is not a priori included in equation (2) and many dielectric systems do not display it, having only a relaxation contribution.

It will be shown in part III and IV below that extrinsic free charges conductivity may interact with and even build strong dielectric relaxation.

### III Polaron contribution to the dielectric permittivity of perovskites

In this section, we show that residual defects in $ABO_3$ perovskite may lead to extrinsic losses at very low temperatures. Such losses will be discussed in terms of polarons formation at e.g Fe centers.

Polarons in perovskite have been attracting a lot of interest in the recent years. This was mainly triggered by the occurrence of strong magneto-resistance in manganites. Since this magnetic coupling is largely due to orbital ordering among the well ordered lattices of $Mn^{3+}/Mn^{4+}$ cations all sitting in the octahedral site of the $ABO_3$ unit cell. Orbital ordering, charge transfer and lattice distortions are the key ingredients for the very special conduction properties of manganites. Probing the structure of polaron lattices is thus a way to access the free charges/ionic lattice distortion and thus to give a clue for the understanding of magneto-resistance. In these cases, we can say that the polaron density is the same as the material density which is about $10^{23}$ cm$^{-3}$. We will however focus here on much diluted cases with polaron densities much less than $10^{20}$ cm$^{-3}$ and at best about $10^{17}$ cm$^{-3}$. Even if these densities are still high as compared to residual impurities concentration in semiconductor materials, they are a very common feature of oxide perovskites. For example, it took several decades of hard work before reaching the bottom concentration of 1 atomic ppm (about $10^{17}$ cm$^{-3}$) unwanted Iron in $BaTiO_3$ single crystals which is the most studied ferroelectric perovskite [33,Rytz].

The usual ways to probe polaron formation in materials are macroscopic conductivity, thermocurrents or infrared absorptions. The latter has been highly debated in recent literature in the case of $SrTiO_3$ mostly because the effective polaron mass resulting from these experiments display strong fluctuations [11,Scott], [12,Eagles], [4,Waser],[34,Mechelen]. On the other hand, bulk conductivity and thermocurrents [35,Kolodziahni] are not easily access because of the very low level of impurity centres leading to the free charges and to the localisation of polarons. In Fe doped $LiNbO_3$, careful drift mobility measurement 13,[Michel-Calendini] were able to confirm the formation of polarons [36,Zylberstein] at Iron sites. In addition, dynamical conductivity experiments based on the Eliott model 14,[Eliott] were useful in KTaO$_3$:Li and SrTiO$_3$:Ca [37,Levtsik,2002]. As shown in other chapters of this book, ESR 10,[Schirmer] and models [38,Vikhnin] are very useful to help solving these issues. In the remaining of this section, we will restrict the discussion to low temperature dielectric dispersion evidencing thermally activated dielectric losses which we interpreted in terms of polarons [39,Bidault,1995]. The first evidence for extremely low dielectric losses was found by Salce et al in very high purity KTaO$_3$ single crystals at about 30-40K [40,Salce] (figure 3). Since no particular dielectric anomaly is expected in such compound in this temperature range, one had to look for extra mechanism for these losses. Because of their extremely low amplitude, these losses could not be fitted to the dielectric relaxation equation (1). However, Arrhenius plotting of the dielectric losses maximum is still possible and activation energy of about 100meV was found. A step towards understanding the physical origin of these losses was taken when Bidault et al [39,Bidault 1995] could successfully show that such losses are by no way restricted to KTaO$_3$ single crystals since they were observed in more than 50 perovskite samples of different composition, morphology and ferroelectric properties (standard ferroelectrics, quantum paraelectric and quantum ferroelectrics). Even more interesting was the common activation energy of about 100meV which was found in all these compounds and latter confirmed in additional materials [41,Trepakov] and fitted previous data [42,Iguchi]. Such a common trend is shown on figure 4 where the straight Arrhenius law on the left side depicts the frequency-temperature map of the dielectric loss maximum. This activation energy is quite compatible with resistivity and Hall measurements performed in BaTiO$_3$ single crystals [43,Kolodz], [44,Gillot]. This shows that the dielectric relaxation which induces the increased dielectric losses is connected to a motion of free charges. However, free charges alone cannot contribute to dielectric relaxation and electrical energy is to be exchanged between these charges and the dielectric lattice. Polarons are the most natural way for this energy exchange to take place. Polaron sites may be unwanted impurities like residual Iron or Oxygen vacancies. Both of these centres may display several valence states which is a necessary requirement for polaron formation and activation. Niobium substituted at the Ti site of BaTiO$_3$ is another well known polaronic site.

Only SrTiO$_3$ displays smaller activation energy at lower temperatures with a tendency towards saturation at the lowest temperatures [45,Viana], [46,Bianchi], [47,Lemanov], [48,Maglione]. Within the polaron model, such trend towards zero-activation energy polarons is in full agreement with the prediction of Mott [15,Mott and Davies]. Moreover, the high temperature (T>15K) extrapolation of the Arrhenius law fits well with high frequency Brillouin Scattering [49,Hehlen] and ESR experiments [50,Müller]. It is thus evidence that defect induced polaron slowing down may be seen on 10 order of magnitudes in frequency on changing the temperature from 40K to 2K [48,Maglione].

**IV Space charges**

**IV.1 Large scale localisation of free charges**

As recalled above, ferroelectric oxides can be considered as broad band gap semi-conductors as for their conductivity properties. When properly selected, even a small amount of substitutions (e.g.a few ppm of Fe in BaTiO$_3$) is able to induce macroscopic increase of the conductivity because of energy levels in the band gap. The standard ways to generate such defects in titanates have been well documented [2,Smyth], [4,Waser], [6,West]:

$$Fe^{3+} + p \rightleftharpoons \square_{Ti}^{4+}$$
$$Nb^{5+} + \bar{e} \rightleftharpoons \square_{Ti}^{4+}$$

$$La^{3+} + \overline{e} \rightleftarrows \square_{Ba}^{2+}$$

$$\square_O^{2-} + p \rightleftarrows \square_O^{-}$$

$$\square_O^{-} + p \rightleftarrows \square_O$$

In these equations, empty squares are vacancies carrying the same charge as the one of the removed ions and e are electrons and p holes. A given configuration of such defects is established at a given annealing temperature for a fixed oxygen partial pressure [2,Smyth]. Let us state from the very beginning that such balanced equations are much less under control as it can be the case in standard semi-conductors. Mostly because of the key role of oxygen vacancies, point defects in ferroelectric perovskites are hardly tuned to better than a ppm ($10^{17} cm^{-3}$) while in Si or Ga-As technology $10^{14} cm^{-3}$ densities of defects are rather common. Even not competing with semi-conductors, defect induced conductivity has reproducible and systematic consequences on various properties of $ABO_3$ ferroelectric materials. Photorefractivity is one of the cleanest evidence of such doped semi-conductor behaviour in Fe-doped $BaTiO_3$ and Fe-doped $LiNbO_3$ [51,Fridkin], [52,Khuktarev]. Another application is the Positive Temperature Coefficient of the Resistance (PTCR) in donor doped $BaTiO_3$ [53,Heywang] which has lead to commercial products since a long time. In both these cases, the impurity induced charge localisation at interfaces is the key mechanism. In photorefractivity, these interfaces are artificially built-in by the optical interference pattern within a single crystal. In PTCR ceramics, the grain boundaries are the natural lattice breaking places where charge depletion is to occur. In single crystals such depletion can appear at the sample/electrode interface, at twin or domains boundaries.

The most popular model for the PTCR effect in ceramics has been described long ago [53,Heywang], [54,Ihrig]. It is based on a description of grain boundaries as insulating barrier layers separating two semi-conducting grains. The whole dc resistivity jump at the phase transition temperature of the grains is then fully explained by a change in the barrier height because of the structural transition in the grains.

We will now focus on the consequences of such depleted interfaces on the dielectric properties of oxides, starting from ferroelectric perovskites.

The first outcome of conducting/insulating oxide interfaces is an increase of the overall capacitance of ceramics. Because of geometric confinement, the ultrathin dielectric layers at grain boundaries of permittivity ε lead to huge capacitance $C = \varepsilon \frac{S}{e}$ because of a very small thickness e and extended surfaces S. Impedance plots –Z" versus Z' are the right tools evidencing the contribution of interfaces to the overall capacitance (see fig1). As already recalled, these provide the capacitance C of each type of interface together with its the dc resistance. Just alike macroscopic capacitors, the charges stored at the grain boundaries are provided by the conducting electrodes which are the inner grains themselves. The main difference with standard capacitors is that the electrodes are not metals in doped oxides. Wolman et al [3,Wollman] have shown very clearly that charge gradients within the grains near the grain boundary interfaces have significant impact. Using acceptor substitution (Ni, Al, Fe) in $SrTiO_3$ ceramics, the electrical barrier at the interfaces were greatly reduced thus confirming the contribution of electrons to the interfacial space charge.

Up to now, low frequency capacitors and dc resistors have been the main focus. We will now examine the consequences of local capacitors and the related charges depletion onto the dynamical dielectric properties of the ceramics. Indeed, the already recalled impedance analysis does not tell much about the redistribution of free charges among the charged point defects. Because of the local nature of the space charge, this is a more complex problem than in single barrier layer capacitors [16,Sze]. Jonscher [30,Jonscher] has shown that dielectric spectroscopy is very useful in this respect. Indeed, the relevant information on dielectric properties is to be extracted in between the two semi-circles of the impedance plots (figure 1). It is when the sample changes from one RC regime to another one that a dielectric relaxation takes place. As shown in figure 1, this relaxation which is hidden in the impedance plot becomes visible in the $\varepsilon_1$ versus $\varepsilon_2$ plots. As a matter of fact, this why the impedance Nyquist plot should not be confused with the dielectric Cole-Cole plot $\varepsilon_2(\varepsilon_1)$, a mistake which is often made because both data set come out from the same experiments.

In the next part, dielectric spectroscopy is shown to be efficient for probing macroscopic space charges in oxides. First, the case of ferroelectric perovskites will be described before going to more recent materials displaying so-called giant dielectric permittivity.

### IV.2 Macroscopic space charges in ferroelectrics

As any oxide, ferroelectric $ABO_3$ perovskite may display increased conductivity under heterovalent cationic substitution (A or B site) and/or reduction through oxygen vacancy creation. The dc conductivity of these perovskites was shown long ago to follow clearly the oxidation/reduction treatment [1,Chan]. Kröger and Vink diagrams were used to model the minimum of resistivity which appeared for each temperature at a given oxygen partial pressure. Not only the reduction state of the samples changes their resistivity but it can also alter the effective dielectric permittivity. For example, the static dielectric permittivity of reduced $BaTiO_3$ single crystals could be raised to about $10^5$ under blocking electrodes conditions [19,Bidault]. On figure 5, a strong maximum of the dielectric permittivity is evidenced whose amplitude could be tuned under oxidation or reduction, reduced samples having the highest effective giant permittivity. On the same plot, one should note that the ferroelectric 130°C transition which does account for the intrinsic polarisability of $BaTiO_3$ does not depend on the reduction state of the crystal. As already pointed out, understanding the origin of the 700°C dielectric maximum requires isothermal dielectric spectroscopy. Such spectra are reported on figure 6 where $\varepsilon'$ is plotted versus frequency for temperatures in the vicinity of the dielectric maximum at 500°C. For all temperatures, a good fitting using equation (2) is achieved. The low frequency tail is fitted using the conductivity $\dfrac{\sigma}{\omega^s}$ term while the high frequency step is modelled by a Cole-Cole relaxation term. Because of the broad frequency range, the parameters used in both fitted contributions do not interact strongly meaning that the two dielectric dispersions are numerically un-correlated. This means that the physical correlation that will be described now is not of numerical origin. To evidence such interplay between the two parameter sets, the fitted conductivity σ and relaxation time τ are plotted on an Arrhenius scale on figure 7. It is clear that the conductivity and relaxation time are both thermally activated with very similar activation energies. Even the cross over from extrinsic to saturated conductivity which induces a cusp in Arrhenius behaviour appears on both parameters at the same temperature. The links between the dielectric relaxation and the dc conductivity are thus twofold:

-at all temperatures, the dispersion step increases with the conductivity, both being proportional to the reduction state of the sample and thus to the amount of oxygen vacancies

-the thermal activation of the conductivity and of the relaxation time is the same

These common features call for a common origin of the two contributions, namely the generation of free charges in the crystal. More precisely, it is the localisation of these free charges at the electrode/crystal interface which artificially increases the dielectric permittivity and thus its relaxation. Such depletion or space charge has been evidenced in 100 different perovskites (single crystals and ceramics) and has been observed by several different research groups [17,Stumpe], [18,Waser], [19,Bidault], [55,Kuwabara]. Even a quantitative agreement could be found between several group since the Arrhenius law from the frequency domain experiments [19,Bidault] could be extrapolated to the time domain experiments which lead to a relaxation frequency of about $10^{-3}$s at 450K [18,Waser]. In any case, the dielectric maximum which is observed at high temperature has nothing to do with an extra microscopic polarisation mechanism in ferroelectric $ABO_3$ perovskites.

Up to now, we have described space charges in the temperature regime where the charge point defects are created. For example, for temperatures higher than 500K, it is well known that oxygen starts diffusing thus leading to the creation of oxygen vacancies. We will know illustrate the relaxation of space charges in the intermediate temperature range where charged point defects are frozen within the lattice.

### IV.3 Macroscopic space charges in other oxides

Once created at high temperatures, charged point defects are fixed on their lattice sites and only electrons or holes may be exchanged among such sites. We show in this part that such mobile charges may be stored at inner interfaces such as grain boundaries or domain walls or any two dimensional defects (twins, dislocations,...). This means that, in single crystals, one should consider these 2-dimensionnal boundaries in addition to the crystal/electrode interface. Even in non-ferroelectric materials, such accumulation of charges may increase strongly the overall dielectric permittivity.
During the last decade, many materials with so-called "giant" dielectric parameters have been investigated. We will list here some of them: Li and Ti substituted [23,Wu], $CaCu_3Ti_4O_{12}$[21, Subramanian], $LuFe_2O_4$ [22,IKeda], $AFe_{1/2}B_{1/2}O_3$ A=Ba, Sr, Ca; B=Nb, Ta, Sb [24,Raevski]. As shown on figure 8, the qualitative features of these materials are all the same:
   -a huge ($\varepsilon>10^4$) and temperature stable dielectric permittivity in the high temperature range (T>300K)
   -a sharp decrease of this dielectric permittivity on cooling to temperatures lower than 300K, the falling temperature decreasing on decreasing the operating frequency
   -a related maximum of the imaginary part ε" of the dielectric permittivity shifting versus frequency just like the ε' step
All these features are the paradigm of a dielectric relaxation. As already stated several times, getting a closer insight in this relaxation requires isothermal experiments where the operating frequency is swept over broad ranges. Such sweeps are shown for $CaCu_3Ti_4O_{12}$ (unpublished) and in $BaTi_{0.85}(Fe_{1/2}Nb_{1/2})_{0.15}O_3$ [56, Abdelkafia] on figure 9 and 10 respectively. The first striking result which was already clear in the thermal experiments (figure 8) is that the dispersion step between the low and high frequency permittivity is temperature independent. In the temperature scale, this results in the flat high temperature variation of ε. This is the first evidence for an extrinsic origin of the dielectric relaxation. Indeed, if lattice related microscopic dipoles were to be involved in the relaxation, an increase of the dispersion step on cooling should be observed. Even in the simplest Debye type model [31,Coehlo], the dispersion step of an assembly of uncorrelated dipoles reads $\Delta\varepsilon = \dfrac{Np^2}{3k_BT\varepsilon_0}$ where N is the density of dipoles p, $k_B$ the Boltzmann constant and $\varepsilon_0$ the vacuum permittivity. The temperature term at the denominator is nothing but the Curie law for a paraelectric assembly of dipoles leading to a divergence of the dispersion step at low temperatures. Any correlation of the dipoles with their surrounding will induce further temperature terms in the dispersion step. This is a very general rule that correlation and the decrease of thermal fluctuations do change the dispersion step probed by dielectric spectroscopy experiments.

Next, fitting of the data using the Jonscher equation 2 [30,Jonscher], lead to a link between the relaxation time activation energy and the inner grain conductivity. In CCTO, the impedance plots [57,West] gave a grain conductivity activation energy of 0.08eV while the dielectric relaxation time also follows an Arrhenius law of activation 0.09eV (figure11)[58,Lunkenheimer]. If, on the same figure 11, one plots the overall activation energy as deduced from the fits of figure 9, one get 0.7eV. This is strong evidence that the space charge that lead to the effective giant dielectric permittivity is not related to an electrode charge localisation or to a macroscopic Maxell-Wagner effect but it is rather limited to the inside of the grains. The picture is now clear: grain boundary acts as very thin capacitors able to store the charges which are provided by the inner grain which are like the conducting electrodes of such capacitors. The Internal Barrier Layer model drawn by West and co-workers is thus the static version of the space charge relaxation picture. Impedance plots were used to quantify the equivalent circuit of CCTO ceramics and dielectric spectroscopy is able to link these to the dispersion of the giant permittivity through the Jonscher law. The same link was successfully established in $BaTi_{0.85}(Fe_{1/2}Nb_{1/2})_{0.15}O_3$ [56,Abdfelkafia] and the same analysis could be applied to $LuFe_2O_4$, NiO and other giant permittivity materials displaying the dielectric behaviour seen on figure 8.

Aside from the artificially high dielectric permittivity, space charges may lead to spurious effects:

-pyroelectric currents were observed in $LuFe_2O_4$[22,Ikeda] and $CaCu_3Ti_4O_{12}$[59,Prakash]. However, the link between these thermally activated pyroelectric currents and the conductivity shows that they should not be confused with sample depolarisation under heating [60, Maglione]. It is rather the space charge delocalisation which leads to an increase of the current as the temperature is raised

-slim hysteresis loops with no sign of saturation [61,Liu]. Because of the non-linear behaviour of the space charges [62, Li], slight opening of electric loops can be observed. Closed loops can easily be recovered under careful compensation of the sample non-linearities

- relaxor behaviour [61,Liu], [63,Ke]: this requires an increase of the dielectric permittivity prior to the start of the relaxation close to the dielectric maximum. It is thus to be pointed out that the dielectric relaxation observed in the four systems of figure 8 has not the usual features of relaxors as some authors did claim. Again, correlation among dipoles is lacking in these space charge systems.

-artificial magnetocapacitance coupling which is all the most observed when the operating frequency is set close to the temperature/frequency range of the relaxation of ε. This picture which was modelled by Catalan [64,Catalan] can be used also for many multiferroic candidates like $BiFeO_3$ [65,Kamba]

### IV.4 Origin of the space charges

In both ferroelectrics at high temperatures (§IV.2) and "giant" permittivity materials in the vicinity of room temperature (§IV.3), space charges are the driving source for strong effective dielectric permittivity and its relaxation. As shown in both kinds of materials, no lattice electric dipoles can be found which could explain permittivities higher than $10^5$ or even $10^6$. For example, in $BaTiO_3$ single crystals, the high temperature space charge may be tuned under reduction while the 130°C dielectric maximum at the ferroelectric phase transition is insensitive to the oxidation state (figure 5). The difference between pure ferroelectrics and "giant" permittivity materials is that the former require high temperature and well defined annealing atmospheres for the space charge to be induced. As a consequence, the space charge relaxation displays large activation energies. On the other hand, "giant" permittivity materials at room temperature undergo a space charge relaxation at temperatures lower than about 300K. In the same time, the activation energy of this relaxation is less than 100meV which do not fit the macroscopic conductivity activation but does agree with the inner grain conductivity activation. The explanation for these various activation energies is to be found in the conductivity which is the source of space charge recombination between interfaces. In pure ferroelectrics, the only conductivity is the intrinsic excitation of electrons across the band gap and/or the motion of oxygen. Both these charge motions need high temperature to be excited and so do the space charge recombination. In giant dielectric materials, electrons or holes may flow among charged point defects within a grain:

-in $CaCu_3Ti_4O_{12}$, $Cu^{2+}$ equilibrated with $Cu^+$ and oxygen vacancies can provide the trapping centres for electron hoping. Moreover, it has been shown that $Ti^{4+}$ substituted into the Cu sites may increase this conduction mechanism [66,Chen]. In that case, improved processing and sintering were used as to show the effective contribution of grain boundaries in the increase of the effective dielectric permittivity [67, Marchin]

-in $LuFe_2O_4$, the couple $Fe^{3+}/Fe^{2+}$ may play the same role to increase the electron mobility. We point out that this should not be confused with the previous part dealing with Fe related polarons in perovskites. For polarons, a very low density of Fe (<1%) is required while for giant effective permittivity $LuFe_2O_4$, two Fe are present in each unit cell. Thus the density of free electrons exchanged between Fe centres is so high that it can induce collective localisation at interfaces thus building space charges

-in $AFe_{1/2}B_{1/2}O_3$ A=Ba, Sr, Ca; B=Nb, Ta, Sb.., $Fe^{3+}/Fe^{2+}$ balance is also the source of free charges

-in Li and Ti doped NiO, $Ti^{3+}$ may be the acceptor site for the trapping of free charges

This list is by no way exhaustive of all the compounds which may display giant dielectric properties [6,Morrison], [68,Maglione], [69,Guillemet] [70,Mondal]. We can however draw a picture for the building of space charges in these materials. First, the free charges which are jointly localised at interfaces as to form the space charge are electrons and holes. As a result, the activation energy for the motion of such charges is small, of the order of 100meV or less. We note that this was the right range of polarons activation energy which was found at §III. Hoping of free charges among residual Fe or Oxygen vacancy centres was seen at very low temperatures because full localisation at individual ionic sites requires such near suppression of thermal fluctuations. However, when interfaces like grain boundaries are present, the collective depletion of such charges can take place at much higher temperatures, e.g.300K. It remains that the dynamics of such space charges depends on the individual excitation of free charges inside the grains and so do the space charge relaxation between two interfaces on each side of the grain.

Even in the absence of charged defects, potential gradients at ferroelectric/conductor interfaces was computed using ab-initio methods [71,Ghosez]. Such microscopic insight where shown to result from the polarisation drop at the interface, thus giving support to the phenomenological models of polarisation reversal [72,Ishibashi, 1990]

**V Summary**

Localisation of free charges on point defects in oxides may lead to several types of effective dielectric properties depending on the lengthscale of the localisation:

-when electrons localise on residual defects at the unit cell length scale, the formation of polarons lead to tiny but well defined enhancement of dielectric losses. These losses appear at very low temperatures (T<50K) and have a common activation energy of 90meV in many kind of perovskites. The specific feature of $SrTiO_3$ displaying 0eV activation energy in the quantum range (T<10K) has been underlined

-when collective localisation occurs at two dimensional interfaces like grain boundaries or twins, giant effective dielectric permittivity are observed in many materials whatever their structure or chemical composition. The only prerequisite for the observation of such giant permittivities is charged defects which can induce enhanced conductivity within the grains and dielectric layers at the grain boundaries providing the localisation sites. In $CaCu_3Ti_4O_{12}$, it appears that the inner grain conductivity and the dielectric relaxation have the same value of 90meV and they are usually detected in the intermediate temperature range (100K-400K)

-in the temperature and atmosphere ranges where defects like oxygen vacancies are created, macroscopic space charges lead also to effective dielectric permittivities of several $10^5$. In all the ferroelectric perovskites tested at T>500K, the activation energies of the macroscopic conductivity and of the dielectric relaxation are the same thus calling for an electrode related localisation of charges. In such case, space charge as well as Maxell Wagner models may equivalently work.

Whatever the above relaxation phenomena, the standard dielectric frequency window which we have used (100Hz<f<$10^7$Hz) fixes the temperature range where a given relaxation is to be observed. Using other low frequency or high frequency spectroscopies only allow to follow the relaxation at low or high temperatures respectively. We have demonstrated such a link in the case of polaron relaxation as well as macroscopic space charge.

At last, one should say again that whenever a dielectric relaxation displays thermally activated relaxation time without a clear increase of the relaxation step on cooling, one should think about free charge related phenomena. In such cases, lattice dipoles are not the main driving source since correlation among them should lead to increased dispersion step on temperature lowering thus following standard Langevin or Brillouin model. This is the main reason why most of the observations of effective giant or colossal dielectric permittivity in defected oxides is a transcription to the field of dielectrics of the well-kown barrier layer design of supercapacitors. In such components of giant specific capacitance, the term permittivity is never used because it is meaningless. Nevertheless, improved control of interfaces in advanced ceramics may help using space charges in defected oxides for applications in integrated devices [73,M.Maglione].

**Acknowledgements**

It is my pleasure to acknowledge stimulating interactions with many colleagues from the "ferroelectric community". I am indebted to my PhD advisor Dr U.T.Höchli who introduced me deeply to the physics of dielectric relaxation thus helping me to distinguish between intrinsic an effective permittivities. Continuous and long term contributions from one of my first PhD student, Dr O.Bidault is also acknowledged. I also wish to recall the memory of Pr.G.Channussot who introduced me to the physics of polarons.

**Figure Captions**

Figure 1: (a) Nyquist plot for the impedance $Z^*(\omega)=Z'(\omega)-jZ''(\omega)$ of a simple parallel R,C circuit as sketched in the inset. (b) Cole Cole plot of the complex permittivity $\varepsilon^*(\omega)=\varepsilon_1(\omega)-j\varepsilon_2(\omega)$ for the same circuit where C and R a frequency independent. (c)Such a straight line becomes a semi-circle when a dielectric relaxation occurs which makes both the capacitance and resistance frequency dependent following equation (1). (d) in such a dielectric relaxation regime, the Nyquist Z''(Z') plot goes back to a straight line indicating a fixed phase for the impedance which is at strong variance from the original Nyquist semi-circle of figure 1(a). The arrows show the direction of the frequency sweep.

Figure 2: schematic variation of the real and imaginary part of the dielectric permittivity as a function of frequency. One can distinguish the space charge and dipole relaxation in the sub-GHz frequency range from the atom, electronic and nuclear resonances at higher frequencies. From the Kramers Krönig principle, any variation of the real part leads to an increase of the imaginary part.

Figure 3: imaginary part of the dielectric susceptibility of Ultra High Purity $KTaO_3$. Taking into account the real part which is several thousands in the same temperature range, the resulting dielectric losses are about 0.1%. The operating frequency f and the temperature where the loss maximum occur can be plotted on an Arrhenius scale thus leading to activation energy of about 100meV [40.Salce].

Figure 4: Arrhenius plot of the dielectric loss maximum in more than 50 different perovskite samples. The straight line at the left which has a common slope of about 100MeV includes data from pure and La doped $BaTiO_3$, pure and Nb doped $KTaO_3$, pure and La and Cu doped $PbTiO_3$. Whatever their composition, growth type, morphology and ferroelectric properties of the samples fall on this same Arrhenius line [39.Bidault, 41.Trepakov, 42.Iguchi]. Only pure and Ca substituted $SrTiO_3$ does escape from this common trend with a much lower temperature range for the dielectric losses and a bending down pointing at zero energy activation at ultralow temperatures [45.Viana, 46.Bianchi, 39.Bidault, 47.Lemanov]. The linear extrapolation at higher temperatures of this dielectric relaxation fit well the higher frequencies ESR [49,Müller] and Brillouin scattering data in $SrTiO_3$ [50,Hehlen].

Figure 5: dielectric permittivity of BaTiO3 single crystals cut from the same boule as a function of temperature. Note the logarithmic scale for this 100 kHz ε which evidences the strong variations at high temperatures T>500°C. The broad dielectric maximum in this temperature range is highly sensitive to the oxidation state of the crystal while the ferroelectric anomaly at 130°C suffers from no changes. Curve taken from Bidault et al [19,Bidault]

Figure 6: isothermal frequency scans of the real part of the dielectric permittivity in the vicinity of the strong high temperature maximum. The low frequency tail stems from the conductivity contribution while the high frequency step evidences a dielectric relaxation. These two *a priori* independent processes were fitted using the Jonscher equation (2). The best fits are the continuous lines at every temperature.

Figure 7: Arrhenius plot of the dielectric relaxation (left scale) and of the static conductivity (right) scale as deduced from the fitting of figure 6. The clear link between these two parameters calls for a contribution of free charges to the dielectric relaxation.

Figure 8 dielectric permittivity and losses at several frequencies as a function of temperature for (a) NiO,Li,Ti 23.[Wu,2002], (b) $CaCu_3Ti_4O_{12}$ 21.[Subramanian,2000], (c) $LuFe_2O_4$ 22.[Ikeda, 2005] and (d) $BaFe_{1/2}Nb_{1/2}O_3$ 24[Raevski,2003]. The same trends are observed for all these materials whatever their chemical content, their crystalline structure and their morphology (ceramics, single crystals and ceramics)

Figure 9: isothermal dispersion of the dielectric permittivity at temperatures close to the step like temperature anomaly of figure 8b for $CaCu_3Ti_4O_{12}$. The real (a) and imaginary(b) part were fitted using equation (2) accounting for the dielectric relaxation at high frequencies and the low frequency conductivity. The lines are the result of such best fits

Figure 10 isothermal dispersion of the dielectric permittivity at temperatures close to the step like temperature anomaly of figure 8d for $BaTi_{0.8}(Fe_{1/2}Nb_{1/2})_{0.15}O_3$. The real (a) and imaginary(b) part were fitted using equation (2) accounting for the dielectric relaxation at high frequencies and the low frequency conductivity. The lines are the result of such best fits

Figure 11: Arrhenius plot of the dielectric relaxation time of $CaCu_3Ti_4O_{12}$ as deduced from the isothermal fits of figure 9. Whatever the electrode material (gold, silver paste, InGa alloy), this activation energy is 90meV. On the right scale with the same 4-decades scale is plotted the overall conductivity deduced from the same fits. For the conductivity, the activation energy is 0.7eV.

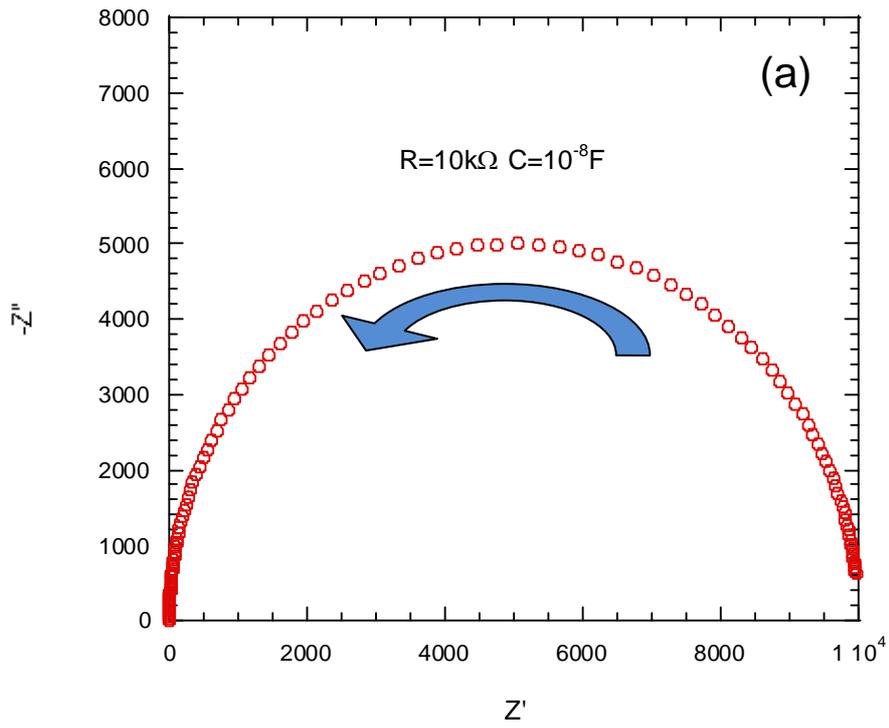

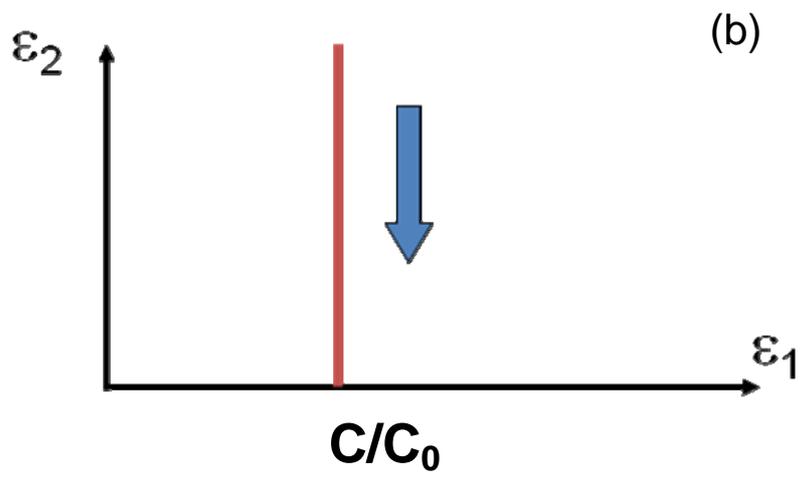

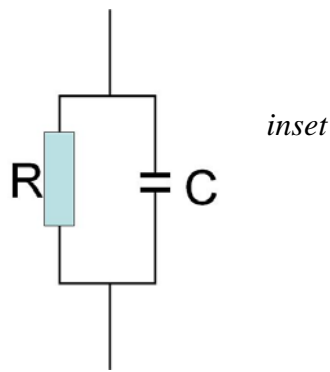

*Fig 1, M.Maglione*

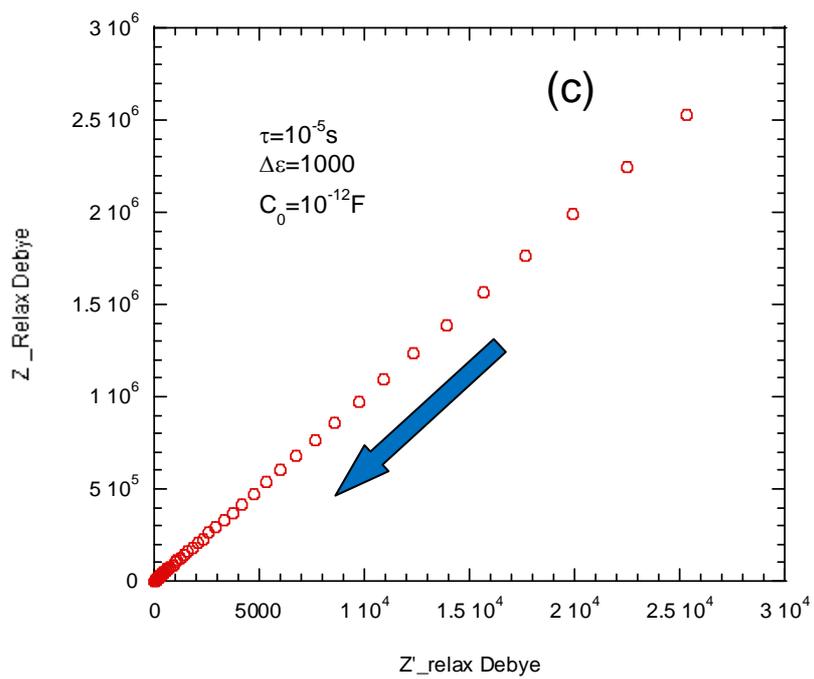

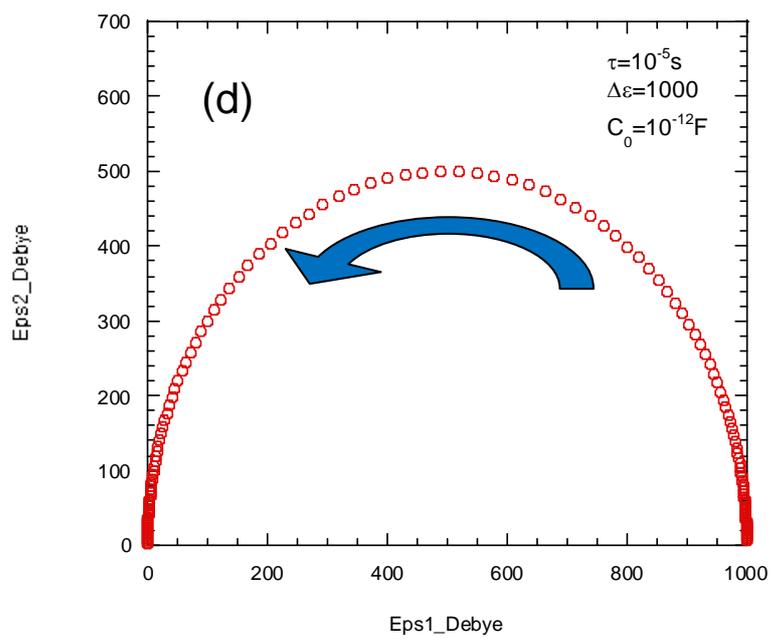

*Fig 1, M.Maglione (cont.)*

*Fig 2 M.Maglione*

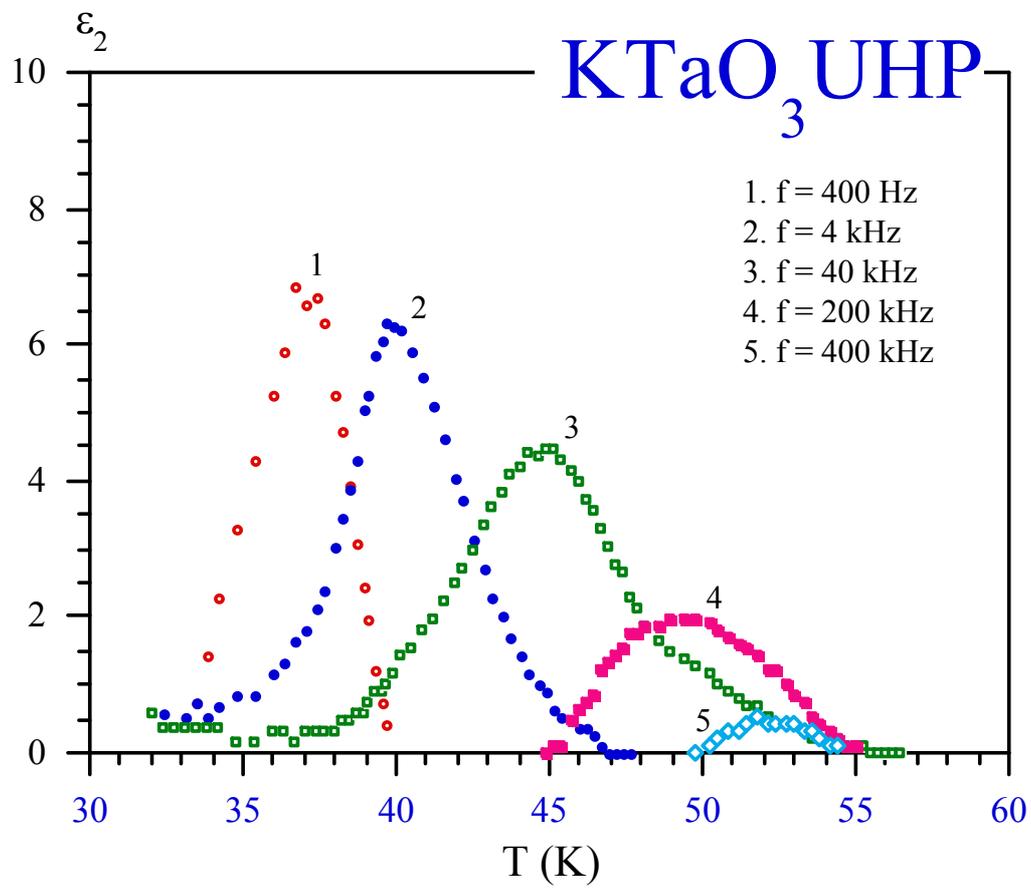

Fig 3, M.Maglione

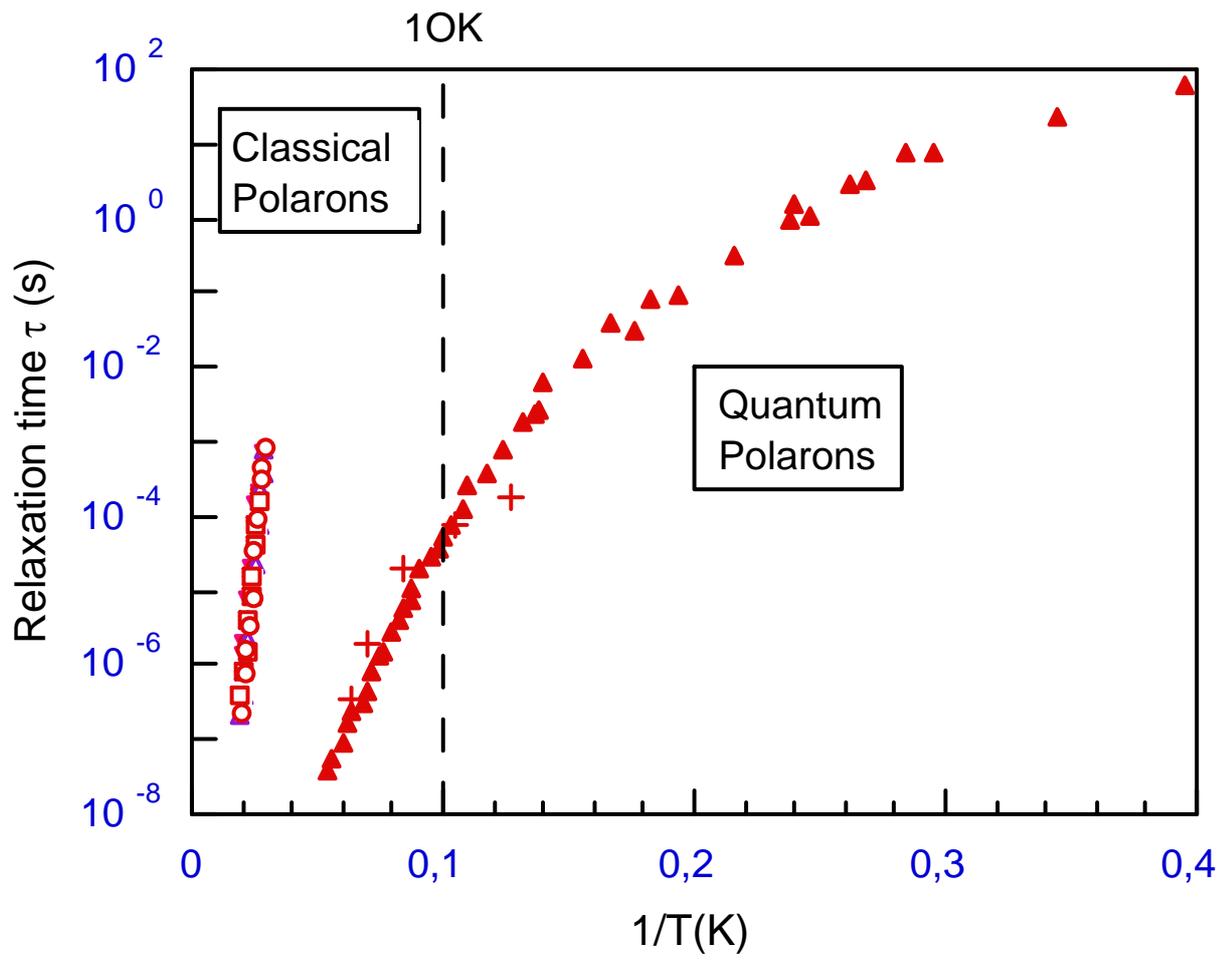

*Fig 4, M.Maglione*

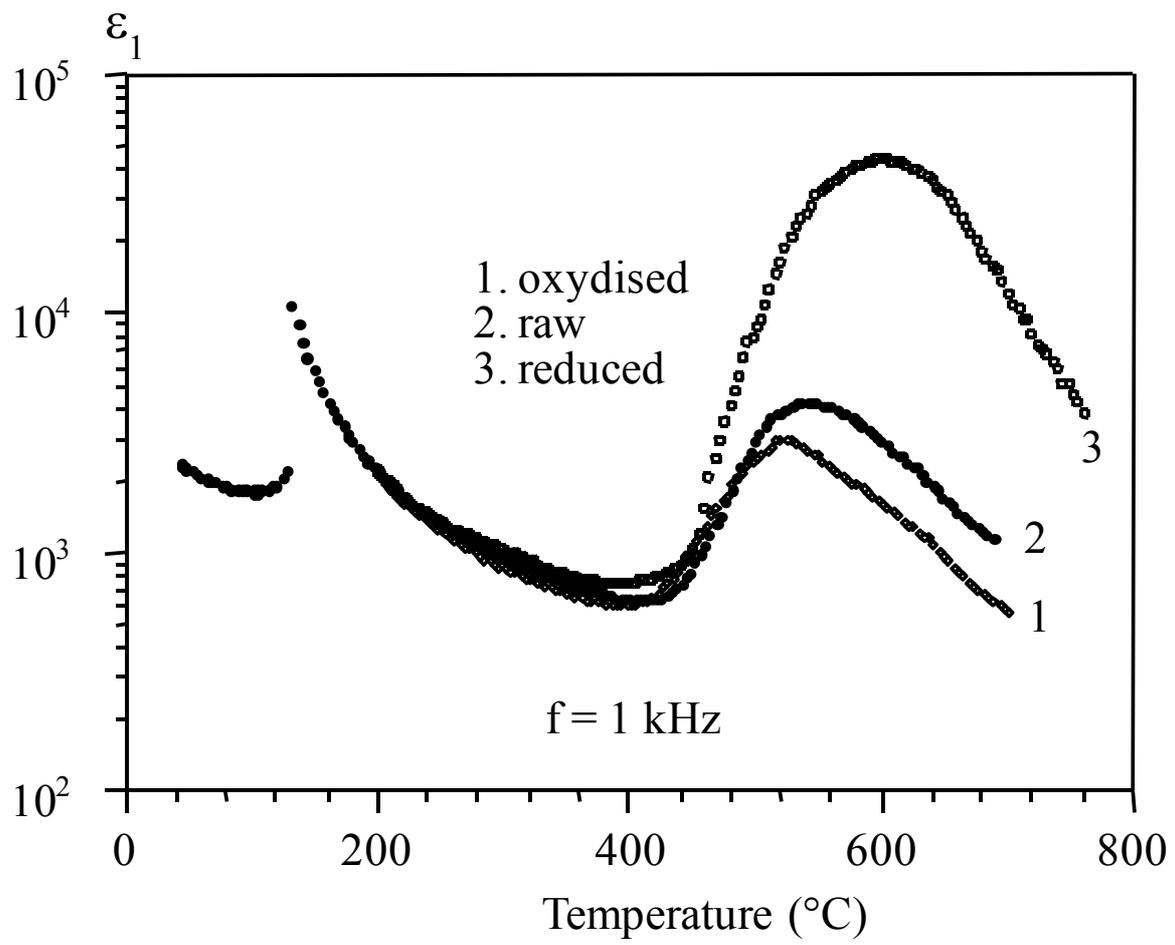

*Fig 5, M.Maglione*

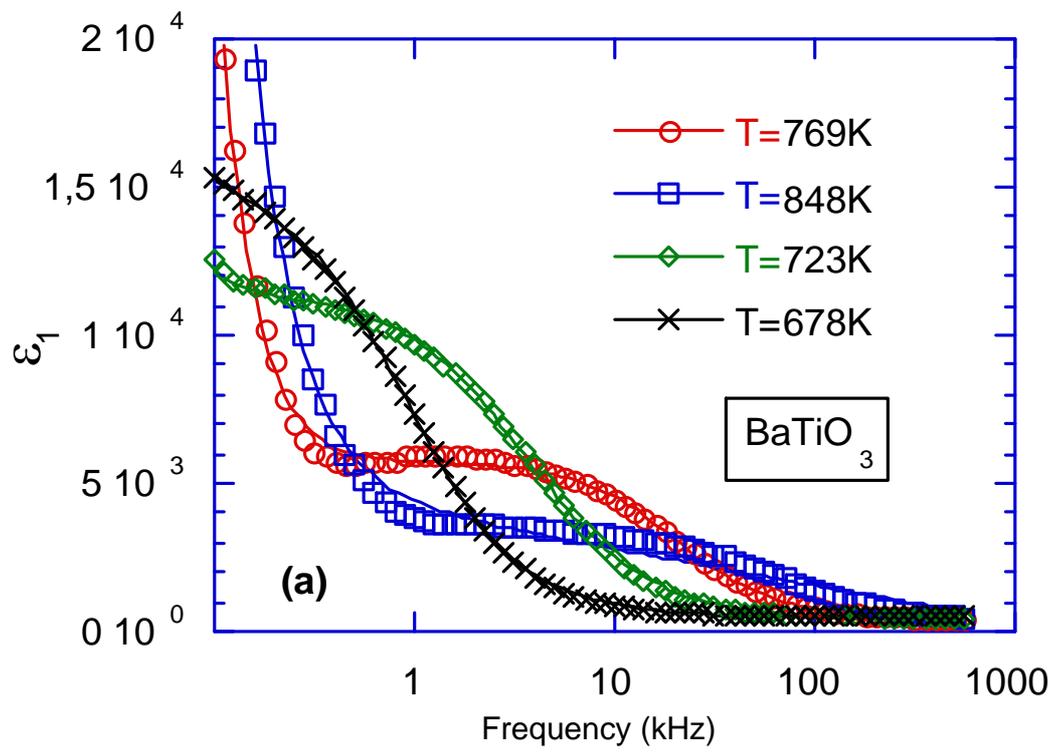

*Fig 6, M.Maglione*

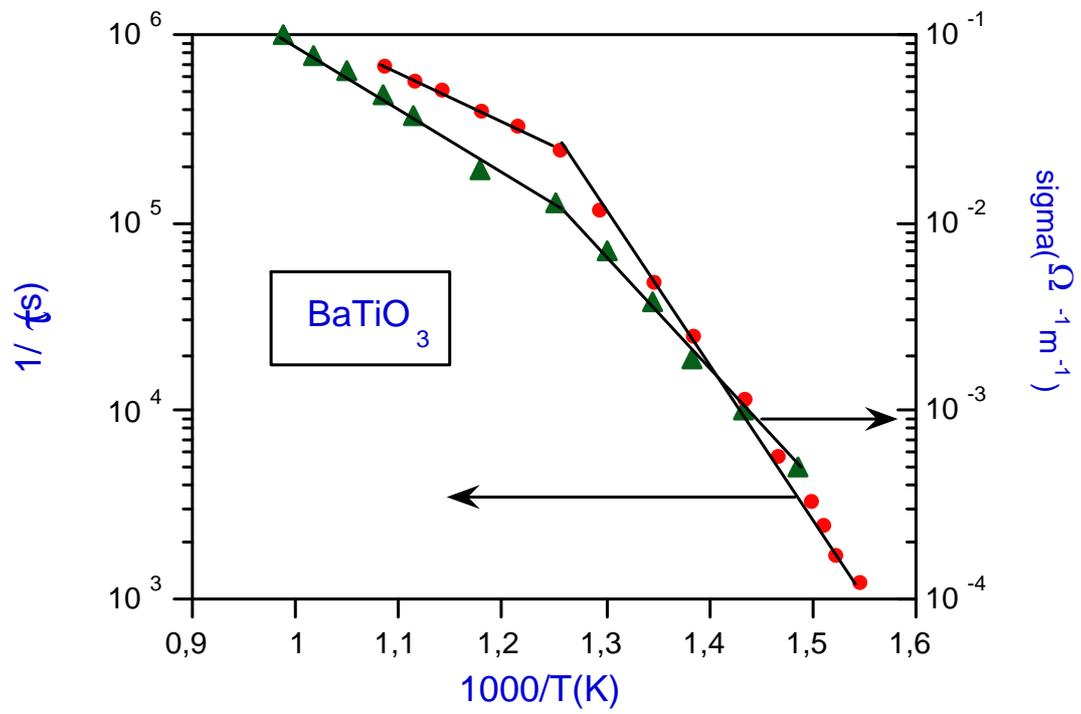

*Fig 7, M.Maglione*

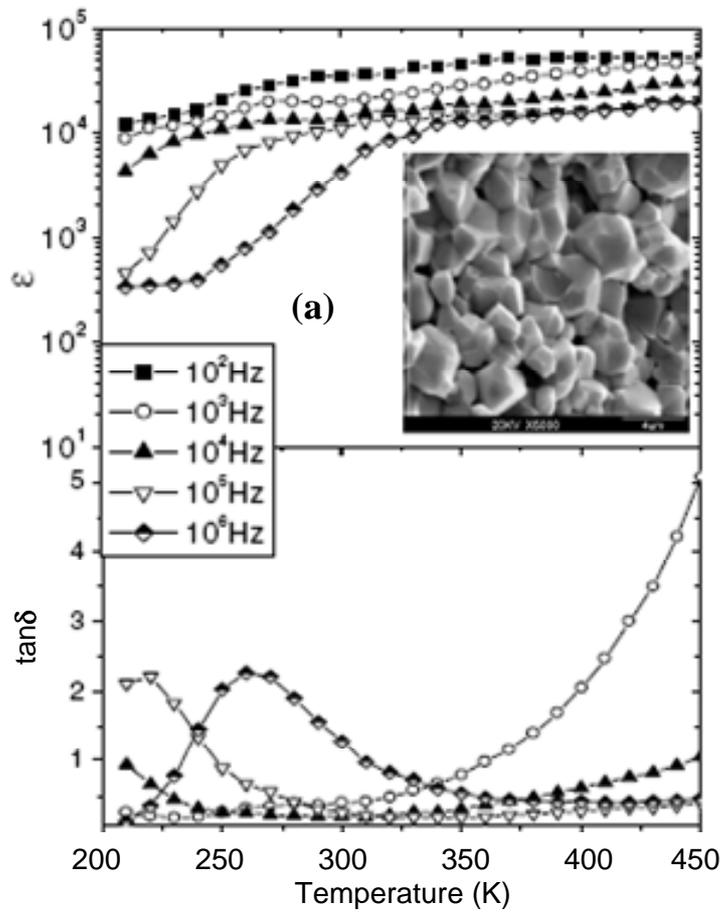

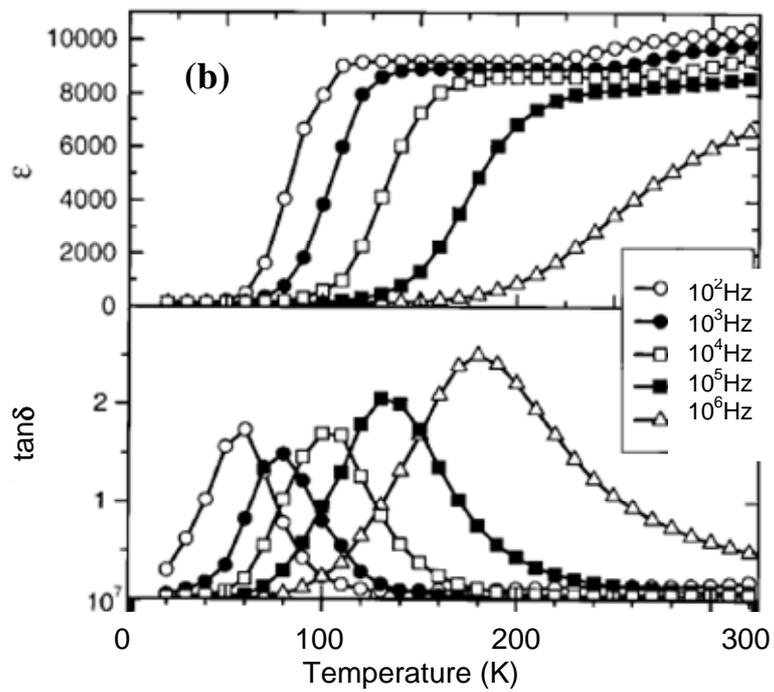

*Fig 8, M.Maglione*

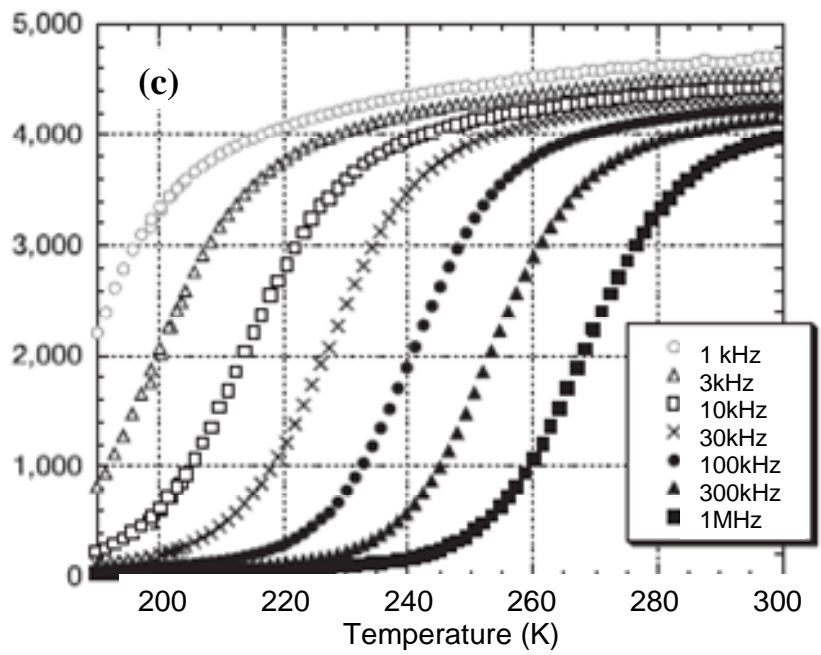

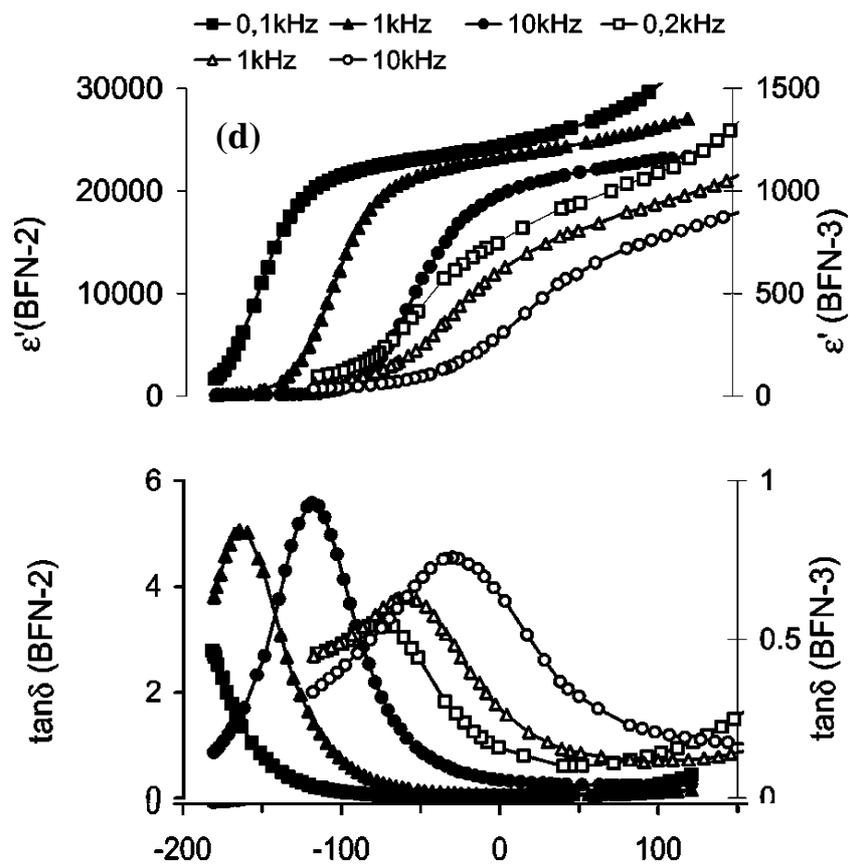

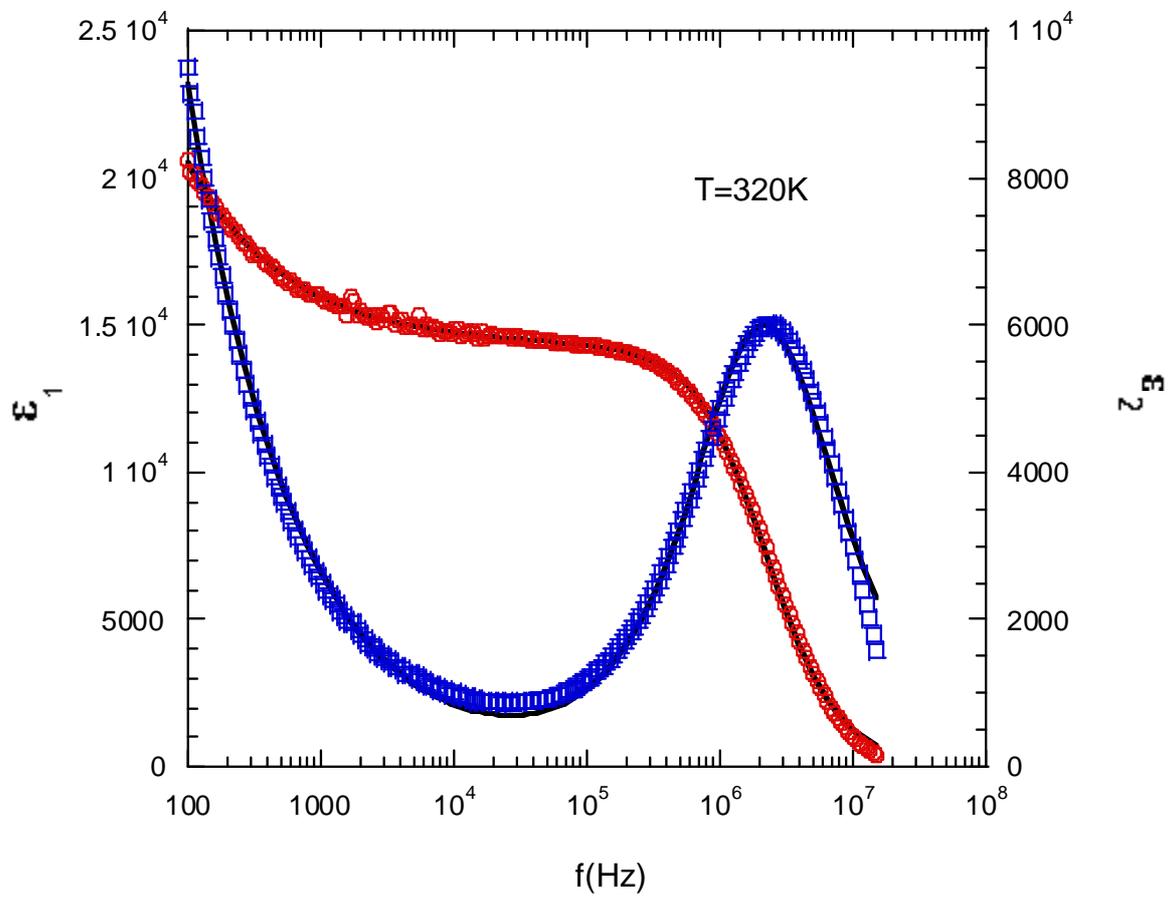

*Fig 9, M.Maglione*

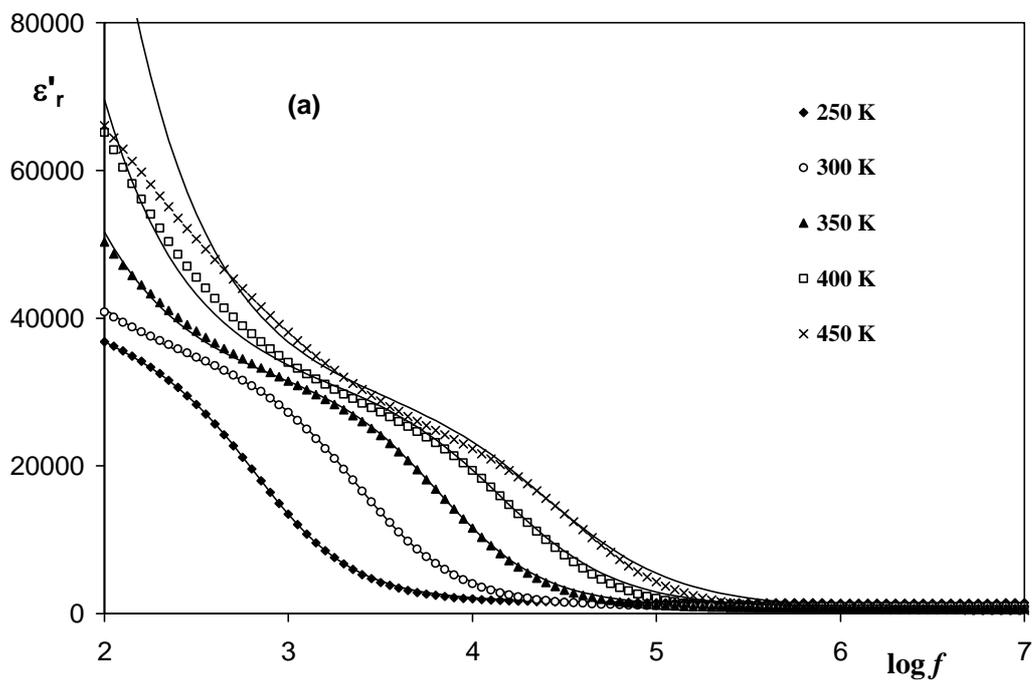

(a)

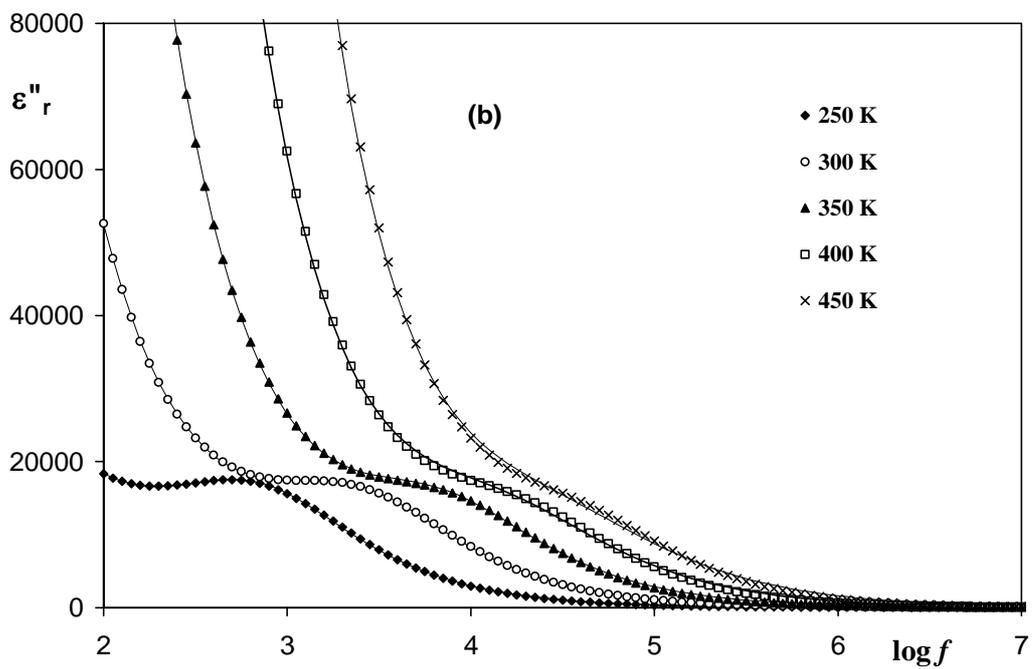

(b)

*Fig 10, M.Maglione*

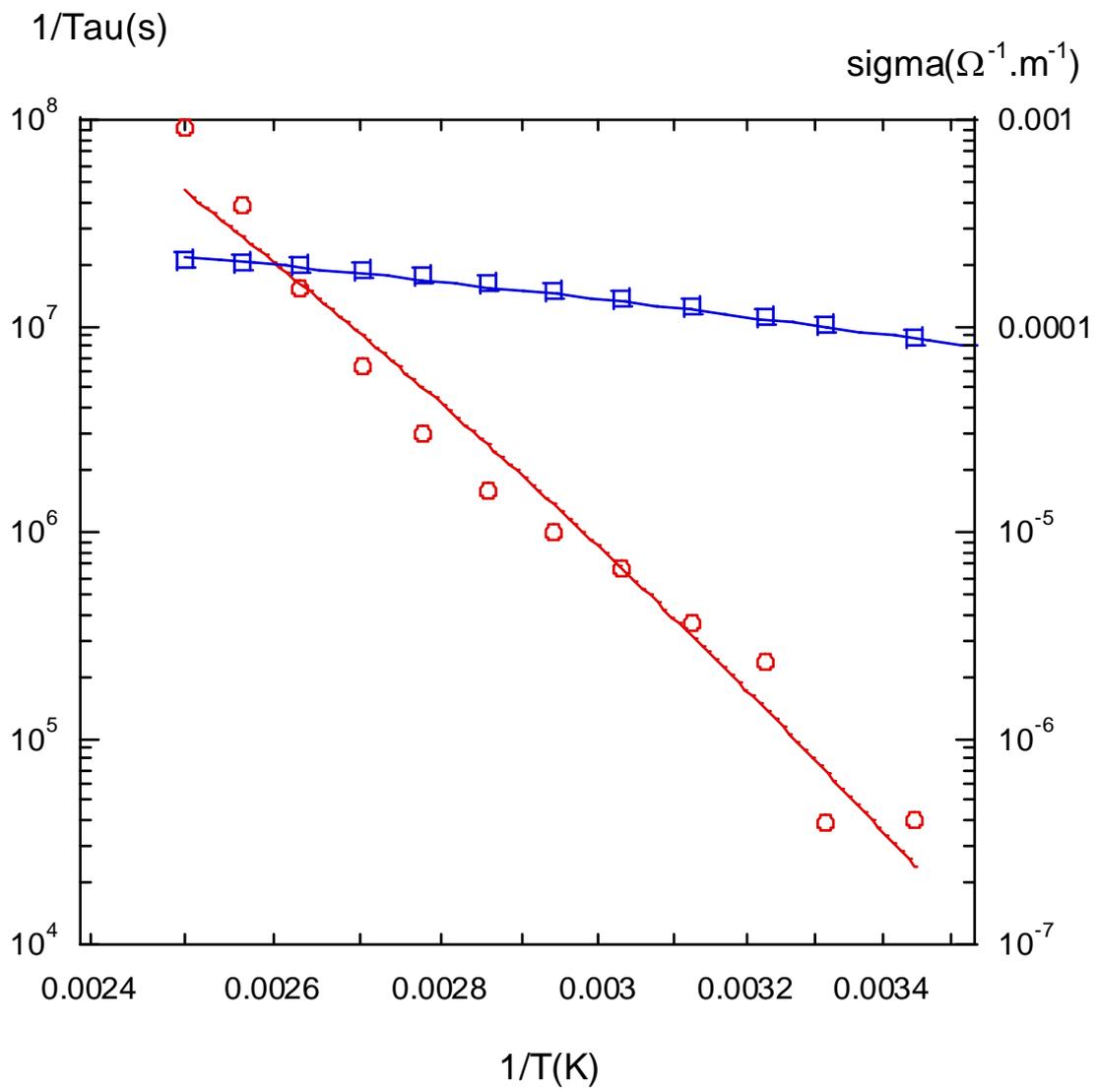

*Fig 11, M.Maglione*